\documentclass[epj]{svjour}
\onecolumn
\usepackage{graphics}
\newcommand{\bm}[1]{\mbox{\boldmath${#1}$}}

\def\inner#1#2{{\bm #1}\cdot {\bm #2}}

\def\Iprot#1#2{P^{#1#2}_{3/2}}
\def\Iproo#1#2{P^{#1#2}_{1/2}}
\def\Sprot#1#2{\Gamma^{#1#2}_{3/2}}
\def\Sproo#1#2{\Gamma^{#1#2}_{1/2}}
\def\half{\frac{1}{2}}
\def\thalf{\frac{3}{2}}
\begin{document}
\title{Bi-local baryon interpolating fields with two flavours}

\author{V. Dmitra\v sinovi\' c\inst{1} \and Hua-Xing Chen\inst{2,3}
}                     
\institute{Institute of Physics, Belgrade University,  Pregrevica
118, Zemun, P.O.Box 57, 11080 Beograd, Serbia, \and Departamento de
F\'{\i}sica Te\'orica and IFIC, Centro Mixto Universidad de
Valencia-CSIC, Institutos de Investigaci\'on de Paterna, Apartado
22085, 46071 Valencia, Spain, \and Department of Physics and State
Key Laboratory of Nuclear Physics and Technology, Peking University,
Beijing 100871, China}
\date{Received: date / Revised version: date}
%
\abstract{ We construct bi-local interpolating field operators for
baryons consisting of three quarks with two flavors, assuming good
isospin symmetry. We use the restrictions following from the Pauli
principle to derive relations/identities among the baryon operators
with identical quantum numbers. Such relations that follow from the
combined spatial, Dirac, color, and isospin Fierz transformations
may be called the (total/complete) Fierz identities. These relations
reduce the number of independent baryon operators with any given
spin and isospin. We also study the Abelian and non-Abelian chiral
transformation properties of these fields and place them into baryon
chiral multiplets. Thus we derive the independent baryon
interpolating fields with given values of spin (Lorentz group
representation), chiral symmetry ($U_L(2) \times U_R(2)$ group
representation) and isospin appropriate for the first angular
excited states of the nucleon.
} 
\maketitle
%
\section{Introduction}
\label{intro}

QCD is at present our best theoretical framework for the description
of hadrons and the chiral symmetry is one of its global symmetries
that plays a key role in hadron physics. Interpolating fields of
hadrons and of baryons in particular have been part and parcel of
lattice QCD and QCD sum rule calculations for almost three decades.
Many such studies suggest that the minimal local structure, i.e.
three quark fields without derivatives, baryon operators
successfully describe properties of the lowest-lying baryon ground
state(s). Moving beyond the ground states to describe even the
lowest-lying excited states turns out to be a challenge in the local
operator approximation, however. Interpolators $B$ for baryons with
spin larger than $3/2$ consisting of three quarks cannot be local
operators in the continuum limit, see Ref.~\cite{Y. Ohnuki 88}.

Indeed, for baryons with total angular momenta larger than $3/2$,
the ``orbital'' angular momentum contribution must be non-zero, that
can only be introduced by means of an additional four-vector, see
Ref.~\cite{Y. Ohnuki 88}. With local operators, there is only one
such four-vector: the four-derivative, or equivalently the
four-momentum of the baryon. Manifestly, an application of the
four-momentum operator to the baryon field cannot change its angular
momentum. One therefore needs another four-vector to ``excite" the
orbital angular momentum, and the two independent separations
between the quarks (the two Jacobi relative coordinate four-vectors)
are precisely what one needs. In plain English, one needs to have at
least two quark fields at two different locations, that leads to a
non-local baryon field.

In this paper we address the question of bi-local baryon fields in
the continuum limit, as the first and simplest extension beyond the
local approximation. The properties of interpolating fields, such as
their Fierz identities~\cite{Ioffe81,Chung:1981cc,Espriu:1983hu} and
chiral properties~\cite{CohenJi,Chen:2008qv,Chen:2009sf} have been
explored in detail only in the lower spin and local operator limit,
whereas the study of higher spin fields has only recently begun, and
that exclusively on the discrete space-time lattice(s), see e.g.
Ref.~\cite{Basak:2005ir}; the higher spin fields in the continuum
space-time have not been dealt with, as yet, to our knowledge.

We shall construct bi-local baryon interpolating fields in such a
way that they belong to reducible representations of the Lorentz
group described by two (half-)integers $(p,q)$ and to irreducible
representations of the isospin $SU(2)$ group, described by the
isospin, a (half-)integer $I$, where the quark field is expressed as
the iso-doublet field. It was not {\it a priori} obvious, however,
that they also belong to the same (irreducible) representations of
the chiral group $SU(2)_R\times SU(2)_L$, where $I_{R,L}$ label the
representations of the right-, and left- isospin groups
$SU(2)_{R,L}$. The ``ordinary'' (vector) isospin $I$ is the quantum
mechanical sum of the right- and left- isospins: $I = |I_{L}-I_{R}|
,\dots , I_{L}+I_{R}$. We have shown in
Ref.~\cite{Nagata:2007di,Nagata:2008xf,Nagata:2008zzc} that the
Fierz identities among local baryon operators also determine their
chiral multiplet structure and shall show here that the same holds
for bi-local fields.

This should not be surprising as the Fierz identities form an
implementation of the Pauli principle, and different permutation
symmetry classes form distinct multiplets of composite particles.
Hence it is necessary to carefully take into account the Fierz
identities also among bi-local baryon operators. The two Jacobi
relative coordinates form the basis of the two-dimensional
irreducible representation of the permutation group $S_3$, which
fact leads to generalized/composite Fierz identities and further
simplifies the classification of the resulting bi-local fields under
the Pauli principle.

The standard isospin formalism greatly facilitates derivation of the
Fierz identities and chiral transformations of baryon operators, due
to the fact that both the quarks and the nucleons belong to the
iso-doublet representation. The composite Fierz identities (i.e. in
the spatial, Dirac, isospin and color space) and the chiral
transformations of baryons are then straightforwardly derived using
the iso-doublet representation.

We give an explicit derivation of these identities for two reasons:
a) this is the first such derivation, to our knowledge; and b)
because of its relative simplicity, we hope that it will show the
way to the chiral $SU(3)_R \times SU(3)_L$ extension, that is
(substantially) more complicated, and encourage others to attack
this and the tri-local field problem. This framework can be applied
to other extensions, such as the inclusion of multi-quark
configurations, and/or of gluon fields into the baryon
interpolators.

This paper is organized as follows. In section~\ref{sec:baryon}, we
firstly define all possible bi-local baryon operators. We classify
the baryon operators according to the representations of the Lorentz
and the isospin groups. Then we apply the Fierz transformation to
obtain Fierz identities among the baryon operators for each
representation of the Lorentz and chiral isospin group. In
section~\ref{sec:chiral_baryon}, we derive the Abelian and
non-Abelian chiral transformations of the baryon operators as
functions of the quarks' chiral transformation parameters, using the
iso-doublet representation. All possible chiral multiplets for the
bi-local baryon operators are displayed by taking into account the
Fierz identities. The final section is a summary and an outlook to
possible future extensions and applications. In
Appendix~\ref{sec:diquarks} we define all possible quark bi-linear
fields with at most one derivative and summarize their chiral
transformations. In Appendix~\ref{sec: app Fierz} we define the
Fierz transformations in the color, flavor and spatial spaces.

\section{Baryon Field Operators}
\label{sec:baryon}

We start with some general comments about three-quark baryon
interpolating operators. An interpolating operator $B$ for baryons
consisting of three quarks cannot be local in general: Indeed, for
baryons with total angular momenta larger than $3/2$, the ``orbital"
angular momentum contribution must be non-zero, and that can only be
described by means of additional four-vector operators. With local
operators, there is only one such four-vector: the four-derivative,
or equivalently the four-momentum of the baryon. Manifestly,
application of this operator to the baryon field cannot change its
angular momentum. One, therefore needs another four-vector to
``excite'' the orbital angular momentum, and the separations between
the quarks (two Jacobi relative coordinates) are precisely what one
needs.

So, in general one must write\footnote{Of course one must include
the color-dependent and path-dependent ``gauge factors". We shall
drop them henceforth, to keep the notation simple.}
\begin{eqnarray}
B(x,y,z) &\sim& \epsilon_{abc} \left(q^T_a(x) \Gamma_1 q_b(y)\right)
\Gamma_2 q_c(z), \label{eq:bgeneral}
\end{eqnarray}
where $q(x)=(u(x),\;d(x))^T$ is an iso-doublet quark field at
location $x$, the superscript $T$ represents the transpose and the
indices $a,\;b$ and $c$ represent the color. Here the antisymmetric
tensor in color space $\epsilon_{abc}$ ensures the baryons' being
color singlets. From now on, we shall omit the color indices always
assuming that the system is a color singlet, which further implies
that any pair of quarks (a ``diquark'') is in a colour anti-triplet
state.

The matrices $\Gamma_{1,2}$ are tensor products of Dirac and isospin
matrices. With a suitable choice of $\Gamma_{1,2}$, the baryon
operators are defined so that they form an irreducible
representation of the Lorentz and isospin groups, as we shall show
in this section.

Note that we use the iso-doublet form for the quark field $q$,
although the explicit expressions in terms of up and down quarks are
usually employed in lattice QCD and QCD sum rule studies. We have
shown in Ref.~\cite{Nagata:2008zzc} that the iso-doublet formulation
leads to a simple classification of baryons into isospin multiplets
and to a straightforward derivation of Fierz identities and chiral
transformations of baryon operators.

As the tri-local fields are substantially more complicated than the
bi-local ones, and neither have been considered in the literature,
as yet, we shall proceed with an analysis of the latter.
\subsection{Bi-local Baryon Fields}
\label{sec:bilocal baryon}

A bi-local interpolating operator $B(x,y)$ for baryons consisting of
three quarks can be generally written as
\begin{eqnarray}
B(x,y) &\sim& \left(q^T(x) \Gamma_1 q(y)\right) \Gamma_2 q(x) +
\left(q^T(x) \Gamma_3 q(x)\right) \Gamma_4 q(y)
\nonumber \\
&=& D^i(x,y) \Gamma^i q(x) + D^j(x,x) \Gamma^j q(y),
\label{eq:bgeneral2}
\end{eqnarray}
where $D^i(x,y)$ are bi-local diquark fields at location $x,y$, see
Appendix~\ref{sec:diquarks}. The Pauli principle relates the two
terms in Eq.~(\ref{eq:bgeneral2}). Here we shall consider the Pauli
principle in two steps.

The first step is to apply the Pauli principle to the first and
second quarks, i.e. to the diquarks, as discussed in
Appendix~\ref{sec:diquarks}. Second, an additional constraint comes
from the permutation of the second and the third quark, which
corresponds to the usual Fierz transformation.

Note that the Fierz transformation connects only baryon operators
belonging to the same Lorentz and isospin group multiplets. We may,
therefore, classify the baryon operators according to their Lorentz
and isospin representations following Chung et
al~\cite{Chung:1981cc}. It has been known that such baryon operators
may couple either to the even- or to the odd-parity states. In the
following discussion all the baryon operators will be defined as
having even parity. We note, however, that two different isospin
baryon operators belonging to the same chiral multiplet may have
opposite parities.

To construct the bi-local baryon fields, we follow the same approach
we used before, and we classify them according to their spin and
isospin. It is convenient to introduce a ``tilde-transposed'' quark
field $\tilde{q}$ as follows
\begin{eqnarray}
\tilde{q}=q^T C\gamma_5 (i\tau_2),
\end{eqnarray}
where $C = i\gamma_2\gamma_0$ is the Dirac field charge conjugation
operator, $\tau_2$ is the second isospin Pauli matrix, whose
elements form the antisymmetric tensor in isodoublet space.

\subsection{$J = {1 \over 2}$ and $I = {1 \over 2}$ fields}

Firstly, we consider the simplest case $D(\frac12,0)_{I=\frac12}$,
where $D(\frac12,0)$ denotes the representation of the Lorentz group
and $I=\frac12$ denotes the isospin. There are twenty bi-local
nucleon operators of $J = {1 \over 2}$ and $I = {1 \over 2}$
\begin{eqnarray}\nonumber
\left. \begin{array}{l}
\left\{ \begin{array}{l} N_1(x,y) = ( \tilde{q}(x) q(y) ) q(x)\, , \\
N_2(x,y) = ( \tilde{q}(x) \gamma_5 q(y) ) \gamma_5 q(x)\, ,  \\
N_3(x,y) = ( \tilde{q}(x) \gamma_\mu q(y) ) \gamma^\mu q(x)\, ,  \\
N_4(x,y) = ( \tilde{q}(x) \gamma_\mu \gamma_5 \tau^i q(y) )
\gamma^\mu \gamma_5 \tau^i q(x)\, ,  \\
N_5(x,y) = ( \tilde{q}(x) \sigma_{\mu\nu} \tau^i q(y) )
\sigma^{\mu\nu} \tau^i q(x)\, , \end{array} \right. \\
\left\{ \begin{array}{l} N_{6}(x,y) = ( \tilde{q}(x) \tau^i q(y) ) \tau^i q(x)\, , \\
N_{7}(x,y) = ( \tilde{q}(x) \gamma_5 \tau^i q(y) ) \gamma_5 \tau^i q(x)\, , \\
N_{8}(x,y) = ( \tilde{q}(x) \gamma_\mu \tau^i q(y) ) \gamma^\mu \tau^i q(x)\, , \\
N_{9}(x,y) = ( \tilde{q}(x) \gamma_\mu \gamma_5 q(y) )
\gamma^\mu \gamma_5 q(x)\, , \\
N_{10}(x,y) = ( \tilde{q}(x) \sigma_{\mu\nu} q(y) ) \sigma^{\mu\nu}
q(x)\, , \end{array} \right.
\end{array} \right. \\
\left. \begin{array}{l} \left\{ \begin{array}{l}
N_{11}(x,y) = ( \tilde{q}(x) q(x) ) q(y)\, , \\
N_{12}(x,y) = ( \tilde{q}(x) \gamma_5 q(x) ) \gamma_5 q(y)\, , \\
N_{13}(x,y) = ( \tilde{q}(x) \gamma_\mu q(x) ) \gamma^\mu q(y)\, , \\
N_{14}(x,y) = ( \tilde{q}(x) \gamma_\mu \gamma_5 \tau^i q(x) )
\gamma^\mu \gamma_5 \tau^i q(y)\, , \\
N_{15}(x,y) = ( \tilde{q}(x) \sigma_{\mu\nu} \tau^i q(x) )
\sigma^{\mu\nu} \tau^i q(y)\, , \end{array} \right. \\
\left\{ \begin{array}{l}
N_{16}(x,y) = ( \tilde{q}(x) \tau^i q(x) ) \tau^i q(y)\, , \\
N_{17}(x,y) = ( \tilde{q}(x) \gamma_5 \tau^i q(x) ) \tau^i \gamma_5 q(y)\, , \\
N_{18}(x,y) = ( \tilde{q}(x) \gamma_\mu \tau^i q(x) ) \gamma^\mu \tau^i q(y)\, , \\
N_{19}(x,y) = ( \tilde{q}(x) \gamma_\mu \gamma_5 q(x) )
\gamma^\mu \gamma_5 q(y)\, , \\
N_{20}(x,y) = ( \tilde{q}(x) \sigma_{\mu\nu} q(x) ) \sigma^{\mu\nu}
q(y)\, . \end{array} \right.
\end{array} \right.
\end{eqnarray}
Among them $N_{16}$-$N_{20}$ vanish due to the Pauli principle.
Using the Fierz identities for the Dirac spin and isospin indices we
obtain the following identities:
\begin{eqnarray}\nonumber
&& \left(\begin{array}{c} N_1 \\ N_2 \\ N_3\\ N_4 \\
N_5 \\ N_{6} \\ N_{7} \\ N_{8} \\ N_{9} \\ N_{10}
\end{array}\right) = \frac18 \left(\begin{array}{cccccccccc}
1 & 1 & 1 & -1 & \frac12 & 1 & 1 & 1 & -1 & \frac12 \\
1 & 1 & -1 & 1 & \frac12 & 1 & 1 & -1 & 1 & \frac12 \\
4 & -4 & -2 & -2 & 0 & 4 & -4 & -2 & -2 & 0 \\
-12 & 12 & -6 & 2 & 0 & 4 & -4 & 2 & -6 & 0 \\
36 & 36 & 0 & 0 & 2 & -12 & -12 & 0 & 0 & -6 \\
3 & 3 & 3 & 1 & -\frac12 & -1 & -1 & -1 & -3 & \frac32 \\
3 & 3 & -3 & -1 & -\frac12 & -1 & -1 & 1 & 3 & \frac32 \\
12 & -12 & -6 & 2 & 0 & -4 & 4 & 2 & -6 & 0 \\
-4 & 4 & -2 & -2 & 0 & -4 & 4 & -2 & -2 & 0 \\
12 & 12 & 0 & 0 & -2 & 12 & 12 & 0 & 0 & -2
\end{array}
\right)\left(\begin{array}{c} N_{11} \\ N_{12} \\ N_{13} \\ N_{14} \\
N_{15} \\ N_{16} \\ N_{17} \\ N_{18} \\ N_{19} \\
N_{20}
\end{array}\right)\, ,
\end{eqnarray}
and
\begin{eqnarray}\nonumber
&& \left(\begin{array}{c} N_{11} \\ N_{12} \\ N_{13} \\ N_{14} \\
N_{15} \\ N_{16} \\ N_{17} \\ N_{18} \\ N_{19} \\
N_{20} \end{array}\right) = \frac18 \left(\begin{array}{cccccccccc}
1 & 1 & 1 & -1 & \frac12 & 1 & 1 & 1 & -1 & \frac12 \\
1 & 1 & -1 & 1 & \frac12 & 1 & 1 & -1 & 1 & \frac12 \\
4 & -4 & -2 & -2 & 0 & 4 & -4 & -2 & -2 & 0 \\
-12 & 12 & -6 & 2 & 0 & 4 & -4 & 2 & -6 & 0 \\
36 & 36 & 0 & 0 & 2 & -12 & -12 & 0 & 0 & -6 \\
3 & 3 & 3 & 1 & -\frac12 & -1 & -1 & -1 & -3 & \frac32 \\
3 & 3 & -3 & -1 & -\frac12 & -1 & -1 & 1 & 3 & \frac32 \\
12 & -12 & -6 & 2 & 0 & -4 & 4 & 2 & -6 & 0 \\
-4 & 4 & -2 & -2 & 0 & -4 & 4 & -2 & -2 & 0 \\
12 & 12 & 0 & 0 & -2 & 12 & 12 & 0 & 0 & -2
\end{array}
\right)\left(\begin{array}{c} N_1 \\ N_2 \\ N_3\\ N_4 \\
N_5 \\ N_{6} \\ N_{7} \\ N_{8} \\ N_{9} \\ N_{10}
\end{array}\right)\, .
\end{eqnarray}
Solving these equations, we obtain the following solutions
\begin{eqnarray}
\left(\begin{array}{c} N_6 \\ N_7 \\ N_8 \\ N_9 \\
N_{10} \\ N_{11} \\ N_{12} \\ N_{13} \\ N_{14} \\ N_{15}
\end{array}\right) =
\frac18\left(\begin{array}{ccccc}
-6 & -6 & -6 & -2 & 1 \\
-6 & -6 & 6 & 2 & 1 \\
-24 & 24 & 12 & -4 & 0 \\
8 & -8 & 4 & 4 & 0 \\
-24 & -24 & 0 & 0 & 4 \\
-6 & 2 & 2 & -2 & 1 \\
2 & -6 & -2 & 2 & 1 \\
8 & -8 & -12 & -4 & 0 \\
-24 & 24 & -12 & -4 & 0 \\
72 & 72 & 0 & 0 & -4
\end{array}
\right)\left(\begin{array}{c} N_1 \\ N_2 \\ N_3\\ N_4 \\ N_5
\end{array}\right)\, .
\end{eqnarray}
Therefore, only these five $\left( N_1, N_2, N_3, N_4, N_5 \right)$
of the original twenty operators survive the Pauli principle.

\subsection{$J = {1 \over 2}$ and $I = {3 \over 2}$ fields}

Next we consider $D(\frac12,0)_{I=\frac32}$ fields. Baryon operators
with $I=\frac32$ must contain either the axial-vector or the tensor
diquark, so there are ten bi-local baryon fields with $J = {1 \over
2}$ and $I = {3 \over 2}$ left
\begin{eqnarray}\nonumber
\left. \begin{array}{l} \left\{\begin{array}{l} \Delta^i_4(x,y) = (
\tilde{q}(x) \gamma_\mu \gamma_5 \tau^j q(y) ) \gamma^\mu \gamma_5
P^{ij}_{3/2} q(x)\, , \\
\Delta^i_5(x,y) = ( \tilde{q}(x) \sigma_{\mu\nu} \tau^j q(y) )
\sigma^{\mu\nu} P^{ij}_{3/2} q(x)\, , \\
\end{array}\right. \\
\left\{\begin{array}{l} \Delta^i_6(x,y) = ( \tilde{q}(x) \tau^j q(y)
) P^{ij}_{3/2} q(x)\, , \\
\Delta^i_7(x,y) = ( \tilde{q}(x) \gamma_5 \tau^j q(y) ) \gamma_5
P^{ij}_{3/2} q(x)\, , \\
\Delta^i_8(x,y) = ( \tilde{q}(x) \gamma_\mu \tau^j q(y) ) \gamma^\mu
P^{ij}_{3/2} q(x)\, ,
\end{array}\right.
\end{array}\right. \nonumber \\
\left. \begin{array}{l} \left\{\begin{array}{l} \Delta^i_{14}(x,y) =
( \tilde{q}(x) \gamma_\mu \gamma_5 \tau^j q(x) ) \gamma^\mu \gamma_5 P^{ij}_{3/2} q(y)\, , \\
\Delta^i_{15}(x,y) = ( \tilde{q}(x) \sigma_{\mu\nu} \tau^j q(x) ) \sigma^{\mu\nu} P^{ij}_{3/2} q(y)\, , \\
\end{array}\right. \\
\left\{\begin{array}{l} \Delta^i_{16}(x,y) = ( \tilde{q}(x)
\tau^j q(x) ) P^{ij}_{3/2} q(y)\, , \\
\Delta^i_{17}(x,y) = ( \tilde{q}(x) \gamma_5 \tau^j q(x) ) \gamma_5 P^{ij}_{3/2} q(y)\, , \\
\Delta^i_{18}(x,y) = ( \tilde{q}(x) \gamma_\mu \tau^j q(x) )
\gamma^\mu P^{ij}_{3/2} q(y)\, .
\end{array}\right.
\end{array}\right.
\end{eqnarray}
Among these operators $\Delta^i_{16}$-$\Delta^i_{18}$ vanishes due
to the Pauli principle. Here $\Iprot{i}{j}$ is the
isospin-projection operator for $I=\frac32$, which is defined,
together with an isospin-projection operator $\Iproo{i}{j}$ for
$I=\frac12$, as
\begin{eqnarray}
\Iprot{i}{j}=\delta^{ij}-\frac13 \tau^i\tau^j,\;
\Iproo{i}{j}=\frac13 \tau^i\tau^j.
\end{eqnarray}
The $I=\frac32$ projection operator satisfies $\tau^i
P^{ij}_{\frac32}=0$, which ensures $\tau^i\Delta^i_{4,5}=0$. Again
the Fierz transformation provides the following relations
\begin{eqnarray}
\left(\begin{array}{c} \Delta^i_6 \\ \Delta^i_7 \\ \Delta^i_8 \\
\Delta^i_4 \\ \Delta^i_5
\end{array}\right) &=&
\frac14\left(\begin{array}{ccccc}
1 & 1 & 1 & -1 & \frac12 \\
1 & 1 & -1 & 1 & \frac12 \\
4 & -4 & -2 & -2 & 0 \\
-4 & 4 & -2 & -2 & 0 \\
12 & 12 & 0 & 0 & -2
\end{array}
\right)\left(\begin{array}{c} \Delta^i_{16} \\ \Delta^i_{17} \\ \Delta^i_{18} \\
\Delta^i_{14} \\ \Delta^i_{15}
\end{array}\right)\, ,
\end{eqnarray}
and
\begin{eqnarray}
\left(\begin{array}{c} \Delta^i_{16} \\ \Delta^i_{17} \\ \Delta^i_{18} \\
\Delta^i_{14} \\ \Delta^i_{15}
\end{array}\right) &=&
\frac14\left(\begin{array}{ccccc}
1 & 1 & 1 & -1 & \frac12 \\
1 & 1 & -1 & 1 & \frac12 \\
4 & -4 & -2 & -2 & 0 \\
-4 & 4 & -2 & -2 & 0 \\
12 & 12 & 0 & 0 & -2
\end{array}
\right)\left(\begin{array}{c} \Delta^i_6 \\ \Delta^i_7 \\ \Delta^i_8 \\
\Delta^i_4 \\ \Delta^i_5
\end{array}\right)\, .
\end{eqnarray}
Solving these equations, we obtain the following solutions
\begin{eqnarray}
\left(\begin{array}{c} \Delta^i_6 \\ \Delta^i_7 \\ \Delta^i_8 \\
\Delta^i_{14} \\ \Delta^i_{15}
\end{array}\right) =
\frac14\left(\begin{array}{cc}
2 & -1 \\
-2 & -1 \\
4 & 0 \\
-8 & 0 \\
0 & -8
\end{array}
\right)\left(\begin{array}{c} \Delta^i_4 \\ \Delta^i_5
\end{array}\right)\, .
\end{eqnarray}
Therefore, only two $(\Delta^i_4, \Delta^i_5)$ of the ten $\Delta$
operators are independent under the Pauli principle.

\subsection{$J = {3 \over 2}$ and $I = {1 \over 2}$ fields}

There are two possible fields/Lorentz group representations with $J
= {3 \over 2}$: 1) the $D(1,\frac12)$ and 2) the $D(\frac32,0)$.

\subsubsection{$D(1,\frac12)$ and $I = {1 \over 2}$}

For $J=\thalf$ fields one of the allowed Lorentz representations is
$D(1,\frac12)$. In this case, baryon operators may contain the
vector and the axial-vector, or the tensor diquark. So we altogether
have twelve bi-local baryon fields
\begin{eqnarray}\nonumber
\left. \begin{array}{l} \left\{ \begin{array}{l} N_{3\mu}(x,y) = (
\tilde{q}(x) \gamma_\nu q(y) ) \Gamma^{\mu\nu}_{3/2} \gamma_5 q(x)\, , \\
N_{4\mu}(x,y) = ( \tilde{q}(x) \gamma_\nu \gamma_5 \tau^i q(y) )
\Gamma^{\mu\nu}_{3/2} \tau^i q(x)\, , \\
N_{5\mu}(x,y) = ( \tilde{q}(x) \sigma_{\alpha\beta} \tau^i q(y) )
\Gamma^{\mu\alpha}_{3/2} \gamma^\beta \gamma_5 \tau^i q(x)\, , \end{array}\right. \\
\left\{ \begin{array}{l} N_{8\mu}(x,y) = ( \tilde{q}(x) \gamma_\nu
\tau^i q(y) ) \Gamma^{\mu\nu}_{3/2} \gamma_5 \tau^i q(x)\, , \\
N_{9\mu}(x,y) = ( \tilde{q}(x) \gamma_\nu \gamma_5 q(y) )
\Gamma^{\mu\nu}_{3/2} q(x)\, , \\
N_{10\mu}(x,y) = ( \tilde{q}(x) \sigma_{\alpha\beta} q(y) )
\Gamma^{\mu\alpha}_{3/2} \gamma^\beta \gamma_5 q(x)\, ,
\end{array}\right.
\end{array}\right. \\ \nonumber
\left. \begin{array}{l} \left\{ \begin{array}{l} N_{13\mu}(x,y) = (
\tilde{q}(x) \gamma_\nu q(x) ) \Gamma^{\mu\nu}_{3/2}
\gamma_5 q(y)\, , \\
N_{14\mu}(x,y) =
( \tilde{q}(x) \gamma_\nu \gamma_5 \tau^i q(x) ) \Gamma^{\mu\nu}_{3/2} \tau^i q(y)\, , \\
N_{15\mu}(x,y) = ( \tilde{q}(x) \sigma_{\alpha\beta} \tau^i q(x)
)\Gamma^{\mu\alpha}_{3/2}
\gamma^\beta \gamma_5 \tau^i q(y)\, , \end{array}\right. \\
\left\{ \begin{array}{l} N_{18\mu}(x,y) = ( \tilde{q}(x) \gamma_\nu
\tau^i q(x) ) \Gamma^{\mu\nu}_{3/2}
\gamma_5 \tau^i q(y)\, , \\
N_{19\mu}(x,y) =
( \tilde{q}(x) \gamma_\nu \gamma_5 q(x) ) \Gamma^{\mu\nu}_{3/2} q(y)\, , \\
N_{20\mu}(x,y) = ( \tilde{q}(x) \sigma_{\alpha\beta} q(x)
)\Gamma^{\mu\alpha}_{3/2} \gamma^\beta \gamma_5 q(y)\, .
\end{array}\right.
\end{array}\right.
\end{eqnarray}
Among them $N_{18\mu}$-$N_{20\mu}$ vanishes due to the Pauli
principle. Similarly to the isospin projection operators,
$\Sprot{\mu}{\nu}$ is the spin-projection operator for $J=\frac32$
states, which is defined, together with the $J=\frac12$ projection
operator $\Sproo{\mu}{\nu}$, by
\begin{eqnarray}
\Sprot{\mu}{\nu}=g^{\mu\nu}-\frac14 \gamma^\mu\gamma^\nu, \;
\Sproo{\mu}{\nu}=\frac14 \gamma^\mu\gamma^\nu. \label{10jan08eq1}
\end{eqnarray}
Owing to this projection operator, the $J=\frac32$ baryon operators
satisfy the Rarita-Schwinger condition $\gamma_\mu N_{3,4,5}^\mu=0$.
The Fierz transformation provides the following relations
\begin{eqnarray}
\left(\begin{array}{c} N_{3\mu} \\ N_{4\mu} \\
N_{5\mu} \\ N_{8\mu} \\ N_{9\mu} \\
N_{10\mu} \end{array}\right) &=& \frac14\left(\begin{array}{cccccc}
1 & 1 & 1 & 1 & 1 & 1 \\
3 & -1 & 1 & -1 & 3 & -3 \\
6 & 2 & 0 & -2 & -6 & 0 \\
3 & -1 & -1 & -1 & 3 & 3 \\
1 & 1 & -1 & 1 & 1 & -1 \\
2 & -2 & 0 & 2 & -2 & 0
\end{array}
\right) \left(\begin{array}{c} N_{13\mu} \\ N_{14\mu} \\
N_{15\mu} \\ N_{18\mu} \\ N_{19\mu} \\
N_{20\mu}
\end{array}\right)\, ,
\end{eqnarray}
and
\begin{eqnarray}
\left(\begin{array}{c} N_{13\mu} \\ N_{14\mu} \\
N_{15\mu} \\ N_{18\mu} \\ N_{19\mu} \\
N_{20\mu} \end{array}\right) &=& \frac14\left(\begin{array}{cccccc}
1 & 1 & 1 & 1 & 1 & 1 \\
3 & -1 & 1 & -1 & 3 & -3 \\
6 & 2 & 0 & -2 & -6 & 0 \\
3 & -1 & -1 & -1 & 3 & 3 \\
1 & 1 & -1 & 1 & 1 & -1 \\
2 & -2 & 0 & 2 & -2 & 0
\end{array}
\right) \left(\begin{array}{c} N_{3\mu} \\ N_{4\mu} \\
N_{5\mu} \\ N_{8\mu} \\ N_{9\mu} \\
N_{10\mu}
\end{array}\right)\, .
\end{eqnarray}
Solving these equations, we obtain the following linear relations
\begin{eqnarray}
\left(\begin{array}{c} N_{8\mu} \\ N_{9\mu} \\ N_{10\mu} \\ N_{13\mu} \\
N_{14\mu} \\ N_{15\mu}
\end{array}\right) =
\frac12\left(\begin{array}{ccc}
-3 & 1 & 1 \\
-1 & -1 & 1 \\
-2 & 2 & 0 \\
-1 & 1 & 1 \\
3 & -3 & 1 \\
6 & 2 & -2
\end{array}
\right)\left(\begin{array}{c} N_{3\mu} \\ N_{4\mu} \\ N_{5\mu}
\end{array}\right)\, .
\end{eqnarray}
Thus, we take $(N_{3\mu}, N_{4\mu}, N_{5\mu})$ as the independent
fields.

\subsubsection{$D(\frac32,0)$ and $I = {1 \over 2}$}

There are four other fields with $J = {3 \over 2}$ and $I = {1 \over
2}$ in the $D(\frac32,0)$ Lorentz group representation, i.e. that
have two Lorentz indices
\begin{eqnarray}
N_{5\mu\nu}(x,y) &=& ( \tilde{q}(x) \sigma_{\alpha\beta} \tau^i q(y)
) \Gamma^{\mu\nu\alpha\beta}_{3/2}
\tau^i q(x)\, , \\
N_{10\mu\nu}(x,y) &=& ( \tilde{q}(x) \sigma_{\alpha\beta} q(y) )
\Gamma^{\mu\nu\alpha\beta}_{3/2} q(x)\, ,
\\
N_{15\mu\nu}(x,y) &=& ( \tilde{q}(x)
\sigma_{\alpha\beta} \tau^i q(x) ) \Gamma^{\mu\nu\alpha\beta}_{3/2} \tau^i q(y)\, , \\
N_{20\mu\nu}(x,y) &=& ( \tilde{q}(x) \sigma_{\alpha\beta} q(x) )
\Gamma^{\mu\nu\alpha\beta}_{3/2} q(y)\, .
\end{eqnarray}
Among them $N_{20\mu\nu}$ vanishes due to the Pauli principle. Here
$\Gamma^{\mu\nu\alpha\beta}$ is another $J=\frac32$ projection
operator defined as
\begin{eqnarray}
\Gamma^{\mu\nu\alpha\beta}=\left(g^{\mu\alpha}g^{\nu\beta} -\half
g^{\nu\beta}\gamma^\mu\gamma^\alpha +\half
g^{\mu\beta}\gamma^\nu\gamma^\alpha +\frac16
\sigma^{\mu\nu}\sigma^{\alpha\beta}\right)\, ,
\end{eqnarray}
The Fierz transformation provides the following relations
\begin{eqnarray}
\left(\begin{array}{c} N_{10\mu\nu} \\ N_{5\mu\nu}
\end{array}\right) = \frac12 \left(\begin{array}{cc}
1 & 1 \\
3 & -1
\end{array}
\right)\left(\begin{array}{c} N_{20\mu\nu} \\ N_{15\mu\nu}
\end{array}\right)\, ,
\end{eqnarray}
and
\begin{eqnarray}
\left(\begin{array}{c} N_{20\mu\nu} \\ N_{15\mu\nu}
\end{array}\right) = \frac12 \left(\begin{array}{cc}
1 & 1 \\
3 & -1
\end{array}
\right)\left(\begin{array}{c} N_{10\mu\nu} \\ N_{5\mu\nu}
\end{array}\right)\, .
\end{eqnarray}
Solving these equations, we obtain 
\begin{eqnarray}
N_{10\mu\nu} = - N_{5\mu\nu} \, , N_{15\mu\nu} = - 2 N_{5\mu\nu} \,
.
\end{eqnarray}
Thus, we take $N_{5\mu\nu}$ as the independent field.

We have, therefore, the grand total of four independent bi-local
baryon fields $(N_{3\mu}, N_{4\mu}, N_{5\mu}, N_{5\mu\nu})$ with $J
= {3 \over 2}$ and $I = {1 \over 2}$.

\subsection{$J = {3 \over 2}$ and $I = {3 \over 2}$ fields}

There are two possible fields/Lorentz group representations with $J
= {3 \over 2}$: 1) the $D(1,\frac12)$ and 2) the $D(\frac32,0)$.

\subsubsection{$D(1,\frac12)$ and $I = {3 \over 2}$}

For $D(1,\frac12)_{I=\frac32}$, there are six operators
\begin{eqnarray}
&& \left\{\begin{array}{l} \Delta^i_{4\mu}(x,y) = ( \tilde{q}(x)
\gamma_\nu
\gamma_5 \tau^j q(y) ) \Gamma^{\mu\nu}_{3/2} P^{ij}_{3/2} q(x)\, , \\
\Delta^i_{5\mu}(x,y) = ( \tilde{q}(x) \sigma_{\alpha\beta} \tau^j
q(y) ) \Gamma^{\mu\alpha}_{3/2} \gamma^\beta \gamma_5 P^{ij}_{3/2}
q(x)\, , \end{array}\right. \\
&& \Delta^i_{8\mu}(x,y) = ( \tilde{q}(x) \gamma_\nu \tau^j q(y) )
\Gamma^{\mu\nu}_{3/2} \gamma_5 P^{ij}_{3/2} q(x)\, ,\\
&& \left\{\begin{array}{l}\Delta^i_{14\mu}(x,y) =
( \tilde{q}(x) \gamma_\nu \gamma_5 \tau^j q(x) ) \Gamma^{\mu\nu}_{3/2} P^{ij}_{3/2} q(y)\, , \\
\Delta^i_{15\mu}(x,y) = ( \tilde{q}(x) \sigma_{\alpha\beta} \tau^j
q(x) )\Gamma^{\mu\alpha}_{3/2}
\gamma^\beta \gamma_5 P^{ij}_{3/2} q(y)\, , \end{array}\right. \\
&& \Delta^i_{18\mu}(x,y) = ( \tilde{q}(x) \gamma_\nu \tau^i q(x) )
\Gamma^{\mu\nu}_{3/2} \gamma_5 P^{ij}_{3/2} q(y)\, .
\end{eqnarray}
Among them $\Delta^i_{18\mu}$ vanishes due to the Pauli principle.
The Fierz transformation provides the following relations
\begin{eqnarray}
\left(\begin{array}{c} \Delta^i_{8\mu} \\
\Delta^i_{4\mu} \\ \Delta^i_{5\mu} \end{array}\right) =
\frac12\left(\begin{array}{cccccc}
1 & 1 & 1 \\
1 & 1 & -1 \\
2 & -2 & 0
\end{array}
\right)\left(\begin{array}{c} \Delta^i_{18\mu} \\
\Delta^i_{14\mu} \\ \Delta^i_{15\mu}
\end{array}\right)\, ,
\end{eqnarray}
and
\begin{eqnarray}
\left(\begin{array}{c} \Delta^i_{18\mu} \\
\Delta^i_{14\mu} \\ \Delta^i_{15\mu} \end{array}\right) =
\frac12\left(\begin{array}{cccccc}
1 & 1 & 1 \\
1 & 1 & -1 \\
2 & -2 & 0
\end{array}
\right)\left(\begin{array}{c} \Delta^i_{8\mu} \\
\Delta^i_{4\mu} \\ \Delta^i_{5\mu}
\end{array}\right)\, .
\end{eqnarray}
Solving these equations, we obtain 
\begin{eqnarray}
\left(\begin{array}{c} \Delta^i_{8\mu} \\ \Delta^i_{14\mu} \\
\Delta^i_{15\mu}
\end{array}\right) =
\left(\begin{array}{cc}
-1 & -1 \\
0 & -1 \\
-2 & -1
\end{array}
\right)\left(\begin{array}{c} \Delta^i_{4\mu} \\ \Delta^i_{5\mu}
\end{array}\right) \, ,
\end{eqnarray}
i.e. there are two independent $\Delta$ fields: $\Delta^i_{4\mu},
\Delta^i_{5\mu}$.

\subsubsection{$D(\frac32,0)$ and $I = {3 \over 2}$}

Finally in the $D(\frac32,0)_{I=\frac32}$ Lorentz representation,
there are only two $\Delta$ operators
\begin{eqnarray}
\Delta^i_{5\mu\nu}(x,y) &=& ( \tilde{q}(x) \sigma_{\alpha\beta}
\tau^j q(y) ) \Gamma^{\mu\nu\alpha\beta}_{3/2} P^{ij}_{3/2} q(x) \,
,
\\ \nonumber \Delta^i_{15\mu\nu}(x,y) &=& (
\tilde{q}(x) \sigma_{\alpha\beta} \tau^j q(x) )
\Gamma^{\mu\nu\alpha\beta}_{3/2} P^{ij}_{3/2} q(y) \, .
\end{eqnarray}
The Fierz transformation provides the following relation
\begin{eqnarray}
\Delta^i_{5\mu\nu} = \Delta^i_{15\mu\nu} \, .
\end{eqnarray}
i.e. there is one independent $\Delta$ field: $\Delta^i_{5\mu\nu}$.

We have, therefore, the grand total of  three independent bi-local
$\Delta$ baryon fields, $(\Delta^i_{4\mu}, \Delta^i_{5\mu},
\Delta^i_{5\mu\nu})$ with $J = {3 \over 2}$ and $I = {3 \over 2}$.

\section{Chiral Transformations}
\label{sec:chiral_baryon}

In this section, we investigate the chiral transformations of
bi-local baryon operators. The chiral mixing of baryon operators is
caused by their diquark components, so it is convenient to classify
the baryon operators according to their diquark chiral multiplets:
$D_1, D_2\in (0,\;0)$, $D_3^\mu , D_4^{\mu i} \in
(\frac12,\;\frac12)$ and $D_5^{\mu\nu i}\in (1,\;0)+(0,\;1)$. The
analysis in the section is similar to our previous
papers~\cite{Nagata:2007di,Nagata:2008zzc} about local fields, so we
simply list the results.

\subsection{$J = {1 \over 2}$}

Under the Abelian chiral transformation the rule, we have
\begin{eqnarray}
\delta_5 N_{1} &=& i a \gamma_5 (N_{1} + 2 N_{2}) \, ,
\\
\delta_5 N_{2} &=& i a \gamma_5 (2 N_{1} + N_{2}) \, ,
\\
\delta_5 N_{3} &=& - i a \gamma_5 N_{3} \, , \\
\delta_5 N_{4} &=& - i a \gamma_5 N_{4} \, , \\
\delta_5 N_{5} &=& 3 a \gamma_5 N_{5} \, ,
\end{eqnarray}
and
\begin{eqnarray}
\delta_5 \Delta^i_{4} &=& - i a \gamma_5 \Delta^i_{4} \, ,
\\
\delta_5 \Delta^i_{5} &=& 3 i a \gamma_5 \Delta^i_{5} \, .
\end{eqnarray}
We can diagonalize $N_{1}$ and $N_{2}$ and obtain
\begin{eqnarray}
\delta_5 (N_{1} + N_{2}) &=& 3 i a \gamma_5 (N_{1} + N_{2}) \, ,
\label{e:nuc+AtrfA} \\
\delta_5 (N_{1} - N_{2}) &=& - i a \gamma_5 (N_{1} - N_{2}) \, .
\label{e:nuc+AtrfB}
\end{eqnarray}

Under the $SU(2)_A$ chiral transformation the rule, we have
\begin{eqnarray}
\delta_5^{\vec{a}} N_{1} &=& i \inner{a}{\tau} \gamma_5 N_{1} \, ,
\\
\delta_5 N_{2} &=& i \inner{a}{\tau} \gamma_5 N_{2} \, ,
\\
\delta_5^{\vec{a}} {N_3} &=& - i \inner{a}{\tau}\gamma_5 {N_3} -
\frac23 i \inner{a}{\tau} \gamma_5 N_4 - 2i \gamma_5 \inner{a}{\Delta_4} \, , \\
\delta_5^{\vec{a}} N_4 &=& - 2 i \inner{a}{\tau}\gamma_5 {N_3} +
\frac13 i \inner{a}{\tau}\gamma_5 N_4 - 2 i
\gamma_5\inner{a}{\Delta_4} \, ,
\\
\delta_{5}^{\vec{a}} N_5 &=& i \inner{a}{\tau} \gamma_5 N_5 \, ,
\end{eqnarray}
and
\begin{eqnarray}
\delta_5^{\vec{a}} \Delta_4^i &=& - 2 i \gamma_5 a^j \Iprot{i}{j}
{N_3} -\frac23 i\gamma_5 a^j P^{ij}_{3/2} N_4 + \frac23 i \tau^i
\gamma_5 \inner{a}{\Delta_4} - i\inner{a}{\tau}\gamma_5 \Delta_4^i
\, ,
\\ \nonumber \delta_5^{\vec{a}} \Delta_5^i &=& - 2 i \tau^i
\gamma_5 \inner{a}{\Delta_5} + 3 i \inner{a}{\tau} \gamma_5
\Delta_5^i \, .
\end{eqnarray}
We find that $N_3$, $N_4$ and $\Delta_4^i$ can be reduced to
irreducible components by taking the antisymmetric linear
combination of the two nucleon fields:
\begin{eqnarray}
\delta_5^{\vec{a}} ({N_3} - N_4) &=& i \inner{a}{\tau} \gamma_5 ({N_3} - N_4) \, ,\\
\delta_5^{\vec{a}}(3 {N_3} + N_4) &=& - i \gamma_5 \left[\frac53
\inner{a}{\tau} (3 {N_3} + N_4) + 8 \inner{a}{\Delta_4} \right] \, ,\\
\delta_5^{\vec{a}} \Delta_4^i &=& - i\gamma_5 \Big[\frac23 a^j
P^{ij}_{3/2} (3 {N_3} + N_4) - \frac23 \tau^i \inner{a}{\Delta_4} +
\inner{a}{\tau} \Delta_4^i \Big] \, .
\end{eqnarray}
Thus we find that $\left(N_1 \pm N_2 \right)$, $({N_3} - N_4)$, and
$N_5$ form four independent $(\frac12,0)$ chiral multiplets,
$\left({N_3} + {1\over3}N_4, \Delta^i_4 \right)$ form one
$(1,\frac12)$ chiral multiplet and $\Delta^i_5$ forms one
$(\frac32,0)$ chiral multiplet.

\subsection{$J = {3 \over 2}$}

Under the Abelian chiral transformation rule, we have
\begin{eqnarray}
\delta_5 N_{3\mu} &=& i a \gamma_5 N_{3\mu} \, ,
\\ \delta_5 N_{4\mu} &=& i a \gamma_5 N_{4\mu} \, ,
\\ \delta_5 N_{5\mu} &=& i a \gamma_5 N_{5\mu} \, ,
\end{eqnarray}
and
\begin{eqnarray}
\delta_5 \Delta^i_{4\mu} &=& i a \gamma_5 \Delta^i_{4\mu} \, ,
\\ \delta_5 \Delta^i_{5\mu} &=& i a \gamma_5 \Delta^i_{5\mu} \, .
\end{eqnarray}

\begin{eqnarray}
\delta_5 N_{5\mu\nu} &=& 3 i a \gamma_5 N_{5\mu\nu} \, ,\\
\delta_5 \Delta^i_{5\mu\nu} &=& 3 i a \gamma_5 \Delta^i_{5\mu\nu} \,
.
\end{eqnarray}

Under the $SU(2)_A$ chiral transformation rule, we have
\begin{eqnarray}
\delta_5^{\vec{a}} N_{3\mu} &=& i \inner{a}{\tau}\gamma_5 N_{3\mu} +
\frac23 i \inner{a}{\tau}\gamma_5
N_{4\mu} + 2i\gamma_5\inner{a}{\Delta_4^\mu} \, , \\
\delta_5^{\vec{a}} N_{4\mu} &=& 2 i \inner{a}{\tau}\gamma_5 N_{3\mu}
- \frac13 i \inner{a}{\tau}\gamma_5
N_{4\mu} + 2i\gamma_5\inner{a}{\Delta_4^\mu} \, , \\
\delta_5^{\vec{a}}N_{5\mu} &=& \frac53 i\inner{\tau}{a}\gamma_5
N_{5\mu} - 4 i \gamma_5\inner{a}{\Delta_{5\mu}} \, , \\
\delta_5^{\vec{a}} \Delta_4^{\mu i}&=& 2 i\gamma_5 a^j
P^{ij}_{\frac32} N_{3\mu} + \frac23 i\gamma_5 a^j P^{ij}_{\frac32}
N_{4\mu} -\frac23 i \tau^i\gamma_5\inner{a}{\Delta_4^\mu}
+i\inner{a}{\tau}\gamma_5
\Delta_4^{\mu i} \, , \\
\delta_5^{\vec{a}} \Delta_5^{\mu i} &=& - \frac43 i\gamma_5 a^j
P^{ij}_{3/2} N_5^\mu - \frac23 i \tau^i\gamma_5
\inner{a}{\Delta_5^\mu} + i\inner{a}{\tau}\gamma_5 \Delta_5^{\mu i}
\, ,
\end{eqnarray}
and
\begin{eqnarray}
\delta_5^{\vec{a}}N_{5\mu\nu} &=& i\inner{\tau}{a}\gamma_5
N_{5\mu\nu} \, ,
\\ \nonumber \delta_5^{\vec{a}} \Delta_{5\mu\nu}^i &=& - 2 i \tau^i
\gamma_5 \inner{a}{\Delta_{5\mu\nu}} + 3 i \inner{a}{\tau} \gamma_5
\Delta_{5\mu\nu}^i \, .
\end{eqnarray}
Thus we find that $(N_{3\mu} - N_{4\mu})$ form one $(\frac12,0)$
chiral multiplet; $\left(N_{3\mu} + \frac13 N_{4\mu},
\Delta^i_{4\mu} \right)$ and $\left(N_{5\mu},
\Delta^i_{5\mu}\right)$ form two independent $(1,\frac12)$ chiral
multiplets; $N_{5\mu\nu} \in (\frac12,0)$, $\Delta^i_{5 \mu\nu} \in
(\frac32,0)$ are also independent chiral multiplets.

\section{Summary and Conclusions}

We have investigated the chiral multiplets consisting of bi-local
three-quark baryon operators, where we took into account the Pauli
principle by way of the Fierz transformation. All spin $\half$ and
$\thalf$ baryon operators were classified according to their Lorentz
and isospin group representations, where spin and isospin projection
operators were employed in Tables~\ref{tab:spin12},
\ref{tab:spin32a}, \ref{tab:spin32b}.
\begin{table}[tbh]
\begin{center}
\caption{The Abelian and the non-Abelian axial charges (+ sign
indicates ``naive", - sign ``mirror" transformation properties) and
the non-Abelian chiral multiplets of $J^{P}=\frac12$, Lorentz
representation $(\frac{1}{2},0)$ nucleon and $\Delta$ fields. All of
the fields are independent and Fierz invariant.}
\begin{tabular}{cccc}
\hline \hline
 & $g_A^{(0)}$ & $g_A^{(1)}$ & $SU_L(2) \times SU_R(2)$ \\
 \hline
 $N_1 - N_2$ & $-1$ & $+1$ & $(\frac12,0) \oplus (0,\frac12)$ \\
 $N_1 + N_2$ & $+3$ & $+1$ & $(\frac12,0) \oplus (0,\frac12)$ \\
 $N_3 - N_4$ & $-1$ & $+1$  & $(\frac12,0) \oplus (0,\frac12)$ \\
 $N_3 + \frac13 N_4$ & $-1$& $-\frac53$ & $(\frac12,1) \oplus (1,\frac12)$ \\
 $\Delta_4$ & $-1$& $-\frac13$ & $(\frac12,1) \oplus (1,\frac12)$ \\
 $N_5$ & $+3$& $+1$ & $(\frac12,0) \oplus (0,\frac12)$ \\
 $\Delta_5$ & $+3$& $+1$ & $(\frac32,0) \oplus (0,\frac32)$ \\
 \hline
\end{tabular}
\label{tab:spin12}
\end{center}
\end{table}
\begin{table}[tbh]
\begin{center}
\caption{The Abelian and the non-Abelian axial charges and the
non-Abelian chiral multiplets of $J^P = \frac{3}{2}$, Lorentz
representation $(1,\frac{1}{2})$ nucleon and $\Delta$ fields. All of
the fields are independent and Fierz invariant.}
\begin{tabular}{cccc}
\hline \hline
 & $g_A^{(0)}$ & $g_A^{(1)}$ & $SU_L(2) \times SU_R(2)$  \\
 \hline
 $N_3^{\mu} - N_4^{\mu}$ & $+1$ & $-1$  & $(0,\frac12) \oplus (\frac12,0)$ \\
 $N_3^{\mu} + \frac13 N_4^{\mu}$ & $+1$& $+\frac53$ & $(1,\frac12) \oplus (\frac12,1)$ \\
 $\Delta_4^{\mu}$ & $+1$& $+\frac13$ & $(1,\frac12) \oplus (\frac12,1)$ \\
 $N_5^{\mu}$ & $+1$ & $+\frac53$ & $(1,\frac12) \oplus (\frac12,1)$ \\
 $\Delta_5^{\mu}$ & $+1$ & $+\frac13$ & $(1,\frac12) \oplus (\frac12,1)$ \\
 \hline
\end{tabular}
\label{tab:spin32a}
\end{center}
\end{table}
\begin{table}[tbh]
\begin{center}
\caption{The Abelian and the non-Abelian axial charges and the
non-Abelian chiral multiplets of $J = \frac{3}{2}$, Lorentz
representation $(\frac{3}{2},0)$ nucleon and $\Delta$ fields. All of
the fields are independent and Fierz invariant.}
\begin{tabular}{cccc}
\hline \hline
& $U_A(1)$ & $SU_A(2)$ & $SU_V(2) \times SU_A(2)$ \\
\hline
$N_{5}^{\mu \nu}$ & $+3$ & $+1$ & $(\frac12,0) \oplus (0,\frac12)$ \\
$\Delta_{5}^{\mu \nu}$ & $+3$ & $+1$ & $(\frac32,0) \oplus (0,\frac32)$ \\
\hline
\end{tabular}
\label{tab:spin32b}
\end{center}
\end{table}

We derived the non-trivial relations among various baryon operators
due to the Fierz transformations, and thus found the independent
baryon fields. Then we found that $\left(N_1 \pm N_2 \right)$,
$({N_3} - N_4)$, and $N_5$ form four independent $(\frac12,0)$
chiral multiplets, whereas $\left({N_3} + \frac13 N_4, \Delta^i_4
\right)$ form one $(1,\frac12)$ chiral multiplet and the independent
field $\Delta^i_5$ also forms a separate $(\frac32,0)$ chiral
multiplet. Thus five nucleons fields, and two $\Delta$s, with
$J=\frac12$ are independent in the bi-local limit, in stark contrast
with the local limit where there are two nucleons and no $\Delta$,
see Ref.~\cite{Nagata:2008zzc}.

In the $J=\frac32$ sector, the $(N_{3\mu} - N_{4\mu})$ form an
independent $(\frac12,0)$ chiral multiplet; $\left(N_{3\mu} +
\frac13 N_{4\mu}, \Delta^i_{4\mu} \right)$ and $\left(N_{5\mu},
\Delta^i_{5\mu} \right)$ form two independent $(1,\frac12)$ chiral
multiplets; $N_{5\mu\nu} \in (\frac12,0)$ and $\Delta^i_{5 \mu\nu}
\in (\frac32,0)$ are also independent chiral multiplets, again in
contrast with the local limit where there is only independent
nucleon field and two independent $\Delta$'s~\cite{Nagata:2008zzc}.

This increase of the number of independent fields is in line with
our expectations from the non-relativistic quark model, where the
number of Pauli-allowed three-quark states in the $L^P = 1^-$ shell
sharply rises from the corresponding number in the ground state.

As in the case of local operators, we showed that the Fierz
transformation connects only the baryon operators with identical
group-theoretical properties, i.e., belonging to the same chiral
multiplet. Then we studied chiral transformations of the bi-local
baryon operators. We found that baryons with different isospins may
mix under the chiral transformations, i.e., they may belong to the
same chiral multiplet. The parity does not play an apparent role in
the chiral properties of the baryon operators at this
(non-dynamical) level.

One of potential applications of our results should be in attempts
to determine the baryons' chiral mixing coefficients/angles, such as
Refs.~\cite{Dmitrasinovic:2009vp,Dmitrasinovic:2009vy,Chen:2009sf,Chen:2010ba}.
This is by no means straightforward business, as there is no
guarantee that these angles are observables. In the case of the
ground state one is fortunate to have the flavor-singlet and octet
axial couplings as an external input into the mixing formalism that
leads to satisfactory fits to baryon/hyperon masses with reasonable
subsequent physical conclusions~\cite{Chen:2009sf,Chen:2010ba}.

The framework presented here holds in standard approaches to QCD,
such as lattice QCD and the QCD sum rules, under the proviso that
chiral symmetry is observed by the approximation used. There is
another (sub-)field where it ought to make an impact: on the class
of fully relativistic approaches, such as those based on the
Bethe-Salpeter equation, to chiral quark
models~\cite{Henriques:1975uh,King:1986wi,Sotiropoulos:1994ub,Stefanis:1999wy,Loring:2001kv,Kvinikhidze:2007qq,Eichmann:2009qa}.

We have employed the standard isospin formalism instead of the
explicit expressions in terms of different flavored quarks in the
flavor components of the baryon fields that are commonplace in this
line of work. By using the isospin formalism, we have been able to
derive all Fierz identities and chiral transformations of the
baryons systematically. The extension to $SU(3)$ is not as
straightforward as one might have imagined, however, so we leave it
for another occasion.

\section*{Acknowledgments}
\label{ack}

We wish to thank Dr. K. Nagata and 
and Prof A. Hosaka, for valuable conversations and correspondence
about the Fierz transformation of triquark fields and (in)dependence
of the nucleon interpolating fields. One of us (V.D.) wishes to
thank Profs. H. Toki and A. Hosaka, for hospitality at RCNP on
several occasion, where this work was started and continued. This
work was financed by the Serbian Ministry of Science and
Technological Development under Grant No.~141025.

\appendix

\section{Bi-local Diquarks}
\label{sec:diquarks}

We begin with bi-linears of two quarks in Eq.~(\ref{eq:bgeneral}).
There are ten non-vanishing possibilities for $\Gamma_1$ with
bi-local fields: besides the familiar five (that survive in the
local approximation limit)
\begin{eqnarray}
D_1(x,y) &=& \tilde{q}(x)q(y),\\
D_2(x,y) &=& \tilde{q}(x)\gamma^5 q(y),\\
D_3^\mu(x,y) &=& \tilde{q}(x)\gamma^\mu q(y),\\
D_4^{\mu i}(x,y) &=& \tilde{q}(x)\gamma^\mu\gamma^5\tau^i q(y),\\
D_5^{\mu\nu i}(x,y) &=& \tilde{q}(x)\sigma^{\mu\nu}\tau^i q(y).
\label{eq:diquark5}
\end{eqnarray}
there are also five new ones (that exist only in the non-local
case):
\begin{eqnarray}
D_6^{i}(x,y) &=& \tilde{q}(x) \tau^i q(y),\\
D_7^{i}(x,y) &=& \tilde{q}(x) \gamma^5 \tau^i q(y),\\
D_8^{\mu i}(x,y) &=& \tilde{q}(x) \gamma^\mu \tau^i q(y),\\
D_9^{\mu}(x,y) &=& \tilde{q}(x) \gamma^\mu\gamma^5 q(y),\\
D_{10}^{\mu\nu}(x,y) &=& \tilde{q}(x) \sigma^{\mu\nu} q(y).
\label{eq:diquark10}
\end{eqnarray}
These quark bi-linear fields, $D_1, D_6^{i}$, $D_2, D_7^{i}$,
$D_3^\mu, D_8^{\mu i}$, $D_4^{\mu i},D_9^{\mu}$ and $D_5^{\mu\nu i},
D_{10}^{\mu\nu}$, shall be referred to as the isoscalar or
isovector, depending on their isospin dependence, scalar,
pseudo-scalar, vector, axial-vector and tensor diquarks, according
to their Lorentz transformation properties~\footnote{Throughout the
present paper, Latin indices $i,\;j$ etc. run over the isospin space
$1,\;2,\;3$, and Greek indices $\alpha,\beta$ etc. run over the
Lorentz space $0,\;1,\;2,\;3$.}.

\subsection{Taylor expansion and the Pauli principle}
\label{ssec:Taylor diq}

Next we make the Taylor expansion and keep terms through the first
order (we define ${\overline{x}}_{\mu} = \frac12 (x+y)_{\mu};
\Delta_{\mu} = (x-y)_{\mu}$):
\begin{eqnarray}
D_1(x,y) &=& \tilde{q}(\overline{x})q(\overline{x}) -
\frac12\Delta^{\alpha}
\tilde{q}(x)(\overrightarrow{\partial}_{\alpha}^{x}-
\overleftarrow{\partial}_{\alpha}^{x}) q(x) + \ldots,\\
D_2(x,y) &=& \tilde{q}(\overline{x})\gamma^5 q(\overline{x}) -
\frac12 \Delta^{\alpha} \tilde{q}(x)\gamma^5
(\overrightarrow{\partial}_{\alpha}^{x}-
\overleftarrow{\partial}_{\alpha}^{x}) q(x) + \ldots ,\\
D_3^\mu(x,y) &=& \tilde{q}(\overline{x})\gamma^\mu q(\overline{x}) -
\frac12 \Delta^{\alpha} \tilde{q}(x)\gamma^\mu
(\overrightarrow{\partial}_{\alpha}^{x}-
\overleftarrow{\partial}_{\alpha}^{x}) q(x) + \ldots,\\
D_4^{\mu i}(x,y) &=& \tilde{q}(\overline{x})\gamma^\mu\gamma^5\tau^i
q(\overline{x}) - \frac12\Delta^{\alpha}
\tilde{q}(x)\gamma^\mu\gamma^5\tau^i
(\overrightarrow{\partial}_{\alpha}^{x}-
\overleftarrow{\partial}_{\alpha}^{x}) q(x) + \ldots,\\
D_5^{\mu\nu i}(x,y) &=& \tilde{q}(\overline{x})\sigma^{\mu\nu}\tau^i
q(\overline{x}) - \frac12\Delta^{\alpha}
\tilde{q}(x)\sigma^{\mu\nu}\tau^i
(\overrightarrow{\partial}_{\alpha}^{x}-
\overleftarrow{\partial}_{\alpha}^{x}) q(x) + \ldots.
\label{eq:diquark5a}
\end{eqnarray}
and
\begin{eqnarray}
D_6^{i}(x,y) &=& \tilde{q}(\overline{x}) \tau^i q(\overline{x}) -
\frac12\Delta^{\alpha}
\tilde{q}(x)(\overrightarrow{\partial}_{\alpha}^{x} -
\overleftarrow{\partial}_{\alpha}^{x}) \tau^i  q(x)+\ldots,\\
D_7^{i}(x,y) &=& \tilde{q}(\overline{x}) \gamma^5 \tau^i
q(\overline{x}) - \frac12\Delta^{\alpha}
\tilde{q}(x)(\overrightarrow{\partial}_{\alpha}^{x} -
\overleftarrow{\partial}_{\alpha}^{x}) \tau^i q(x)+\ldots,\\
D_8^{\mu i}(x,y) &=& \tilde{q}(\overline{x}) \gamma^\mu \tau^i
q(\overline{x}) - \frac12\Delta^{\alpha} \tilde{q}(x)\gamma^\mu
(\overrightarrow{\partial}_{\alpha}^{x} -
\overleftarrow{\partial}_{\alpha}^{x})\tau^i q(x) + \ldots,\\
D_9^{\mu}(x,y) &=& \tilde{q}(\overline{x}) \gamma^\mu\gamma^5
q(\overline{x}) - \frac12\Delta^{\alpha}
\tilde{q}(x)\gamma^\mu\gamma^5
(\overrightarrow{\partial}_{\alpha}^{x} -
\overleftarrow{\partial}_{\alpha}^{x}) q(x) + \ldots,\\
D_{10}^{\mu\nu}(x,y) &=& \tilde{q}(\overline{x}) \sigma^{\mu\nu}
q(\overline{x}) - \frac12\Delta^{\alpha} \tilde{q}(x)\sigma^{\mu\nu}
(\overrightarrow{\partial}_{\alpha}^{x} -
\overleftarrow{\partial}_{\alpha}^{x}) q(x) + \ldots.
\label{eq:diquark10a}
\end{eqnarray}
Taking into account the Pauli principle, we find
\begin{eqnarray}
D_1(x,y) &=& \tilde{q}(\overline{x})q(\overline{x}) + {\cal O}(\Delta^2),\\
D_2(x,y) &=& \tilde{q}(\overline{x})\gamma^5 q(\overline{x}) + {\cal O}(\Delta^2) ,\\
D_3^\mu(x,y) &=& \tilde{q}(\overline{x})\gamma^\mu q(\overline{x}) + {\cal O}(\Delta^2) ,\\
D_4^{\mu i}(x,y) &=& \tilde{q}(\overline{x})\gamma^\mu\gamma^5\tau^i
q(\overline{x})
 + {\cal O}(\Delta^2) ,\\
D_5^{\mu\nu i}(x,y) &=& \tilde{q}(\overline{x})\sigma^{\mu\nu}\tau^i
q(\overline{x}) + {\cal O}(\Delta^2). \label{eq:diquark5b}
\end{eqnarray}
and
\begin{eqnarray}
D_6^{i}(x,y) &=& - \frac12\Delta^{\alpha}
\tilde{q}(x)(\overrightarrow{\partial}_{\alpha}^{x} -
\overleftarrow{\partial}_{\alpha}^{x}) \tau^i  q(x)+\ldots,\\
D_7^{i}(x,y) &=& - \frac12\Delta^{\alpha}
\tilde{q}(x)(\overrightarrow{\partial}_{\alpha}^{x} -
\overleftarrow{\partial}_{\alpha}^{x}) \tau^i q(x)+\ldots,\\
D_8^{\mu i}(x,y) &=& - \frac12\Delta^{\alpha} \tilde{q}(x)\gamma^\mu
(\overrightarrow{\partial}_{\alpha}^{x} -
\overleftarrow{\partial}_{\alpha}^{x})\tau^i q(x) + \ldots,\\
D_9^{\mu}(x,y) &=& - \frac12\Delta^{\alpha}
\tilde{q}(x)\gamma^\mu\gamma^5
(\overrightarrow{\partial}_{\alpha}^{x} -
\overleftarrow{\partial}_{\alpha}^{x}) q(x) + \ldots,\\
D_{10}^{\mu\nu}(x,y) &=& - \frac12\Delta^{\alpha}
\tilde{q}(x)\sigma^{\mu\nu} (\overrightarrow{\partial}_{\alpha}^{x}
- \overleftarrow{\partial}_{\alpha}^{x}) q(x) + \ldots.
\label{eq:diquark10b}
\end{eqnarray}

\subsection{Chiral properties of diquarks}
\label{ssec:chiral diq}

Firstly we look at their Abelian chiral ($U(1)_A$) transformation
properties, which are determined by the $U(1)_A$ transformation of
the quark field,
\begin{eqnarray}
q \to \exp(i\gamma_5 a) q  = q + \delta_5 q ,
\label{e:qU1trf} \\
\tilde{q} \to \tilde{q} \exp(i\gamma_5 a) = \tilde{q} + \delta_5
\tilde{q},
 \label{e:qtildeU1trf}
\end{eqnarray}
where $a$ is an infinitesimal parameter for $U(1)_A$ transformation.
The scalar and pseudo-scalar diquarks transform as
\begin{eqnarray}
\delta_5 D_1 = 2ia D_2,\\
\delta_5 D_2 = 2ia D_1,
\end{eqnarray}
from which it follows that
\begin{eqnarray}
\delta_5 (D_1 - D_2) &=& - 2 i a (D_1 - D_2),\\
\delta_5 (D_1 + D_2) &=& 2 i a (D_1 + D_2),
\end{eqnarray}
the vector and axial-vector diquarks,
\begin{eqnarray}
\delta_5 D_{3,4}= 0,
\end{eqnarray}
the tensor diquark,
\begin{eqnarray}
\delta_5 D_5^{\mu \nu i}= 2 ia D_{11}^{\mu \nu i},
\end{eqnarray}
where ${D}^{\mu\nu i}_{11} = \tilde{q} \sigma^{\mu\nu} \gamma_5
\tau^i q$ is a dual-tensor diquark. Note that baryon operators
constructed from the dual-tensor diquark can be expressed as
functions of the tensor diquark by using the identity
$\sigma^{\mu\nu}\gamma_5 = - \frac{i}{2}
\epsilon^{\mu\nu\alpha\beta} \sigma_{\alpha\beta}$.

Next we consider the axial transformation
\begin{eqnarray}
{q} &\to& \exp (i \gamma_{5} \inner{\tau}{a}){q} = q +
\delta_5^{\vec{a}} q,
\label{eq:Qtrfch}\\
{\tilde q} &\to& {\tilde q}\exp (-i \inner{\tau}{a}\gamma_{5}) =
\tilde{q} + \delta_5^{\vec{a}} \tilde{q}, \label{eq:Qtrf}
\end{eqnarray}%
where $\vec{a}$ is the  triplet for the axial transformation
parameters. It is straightforward to obtain the axial
transformations of the diquarks: for the scalar and pseudo-scalar
diquarks,
\begin{eqnarray}
\delta_5^{\vec{a}} D =0,\; (D=D_1,D_2), \label{eq:chirald1}
\end{eqnarray}%
for the vector and axial-vector diquarks,
\begin{eqnarray}
\delta_5^{\vec{a}} D^\mu_3 = 2i a^i D^{\mu i}_4,
\label{eq:chirald3}\\
\delta_5^{\vec{a}} D^{\mu i}_4 = 2 i a^i D^\mu_3,
\label{eq:chirald4}
\end{eqnarray}%
for the tensor diquark,
\begin{eqnarray}
\delta_5^{\vec{a}} D_5^{\mu\nu i} = - 2 \epsilon^{ijk} a^j
D_{11}^{\mu\nu k}. \label{eq:chirald5}
\end{eqnarray}
These transformations imply that the scalar and pseudo-scalar
diquarks $D_{1,2}$ are chiral scalars (invariants) $(0,\; 0)$. The
vector and axial-vector diquarks $D_{3}^\mu,\; D_4^{\mu i}$ together
belong to the chiral multiplet $(\frac12,\;\frac12)$; therefore they
are chiral partners, similar to the $(\sigma,\;\vec{\pi})$ case. The
tensor diquark transforms into the dual-tensor diquark under
non-Abelian chiral transformations, and the two together form the
chiral multiplet $(1,0)\oplus (0,1)$.

Similarly, the ``Pauli forbidden'' diquarks transform as follows
under the $U_{A}(1)$ axial transformation.

The scalar and pseudo-scalar diquarks transform as
\begin{eqnarray}
\delta_5 D_6 = 2 i a D_7 ,\\
\delta_5 D_7 = 2 i a D_6 ,
\end{eqnarray}
from which follows
\begin{eqnarray}
\delta_5 (D_6 - D_7) &=& - 2 i a (D_6 - D_7),\\
\delta_5 (D_6 + D_7) &=& 2 i a (D_6 + D_7),
\end{eqnarray}
the vector and axial-vector diquarks,
\begin{eqnarray}
\delta_5 D_{8,9}= 0,
\end{eqnarray}
the tensor diquark,
\begin{eqnarray}
\delta_5 D_{10}^{\mu \nu}= 2 ia D_{11}^{\mu \nu},
\end{eqnarray}
where ${D}^{\mu\nu}_{11} = \tilde{q} \sigma^{\mu\nu} \gamma_5 q$ is
a dual-tensor diquark. Note that baryon operators constructed from
the dual-tensor diquark can be expressed as functions of the tensor
diquark by using the identity $\sigma^{\mu\nu}\gamma_5 = -
\frac{i}{2} \epsilon^{\mu\nu\alpha\beta} \sigma_{\alpha\beta}$.

Next we consider the non-Abelian axial transformation,
Eqs.(\ref{eq:Qtrf}). It is straightforward to obtain the axial
transformations of the diquarks: for the scalar and pseudo-scalar
diquarks,
\begin{eqnarray}
\delta_5^{\vec{a}} D_6^i &=& - 2 \epsilon^{ijk} a^j D_7^k,
\nonumber \\
\delta_5^{\vec{a}} D_7^i &=& - 2 \epsilon^{ijk} a^j D_6^k,\;
\label{eq:chirald1b}
\end{eqnarray}%
for the vector and axial-vector diquarks,
\begin{eqnarray}
\delta_5^{\vec{a}} D^\mu_9 = 2 i a^i D^{\mu i}_{8},
\label{eq:chirald3}\\
\delta_5^{\vec{a}} D^{\mu i}_8 = 2 i a^i D^\mu_9,
\label{eq:chirald4}
\end{eqnarray}%
for the tensor diquark,
\begin{eqnarray}
\delta_5^{\vec{a}} D_{10}^{\mu\nu} = 0. \label{eq:chirald5}
\end{eqnarray}
These transformations imply that the scalar and pseudo-scalar
diquarks $D_6^i ,D_7^i$ together form the chiral multiplet
$(1,0)\oplus (0,1)$, they are chiral partners, similar to the
$(\vec{\rho},\;\vec{a_{1}})$ mesons. The vector and axial-vector
diquarks $D_{3}^\mu,\; D_4^{\mu i}$ together belong to the chiral
multiplet $(\frac12,\;\frac12)$; therefore they are chiral partners,
similar to the $(\sigma,\;\vec{\pi})$ case. The isoscalar tensor
diquark transforms into the isoscalar dual-tensor diquark under
non-Abelian chiral transformations, so they are chiral scalars
(invariants) $(0,\; 0)$.

All spin $0$ and $1$ diquark operators were classified according to
their Lorentz and isospin group representations, in Table
\ref{tab:diquarks}.
\begin{table}[tbh]
\begin{center}
\caption{The Abelian $U_A(1)$ and the non-Abelian $SU_L(2) \times
SU_R(2)$ chiral transformation properties/axial charges and the
Lorentz group representations of the diquark fields. In the last
column we show the sign under transposition of the two quarks in the
colour $\overline{3}_{\rm }$ state.}
\begin{tabular}{ccccc}
\hline \hline
& $U_A(1)$ & Lorentz group $SO(3,1)$ & $SU_L(2) \times SU_R(2)$ & exchange \\
\hline
$D_1 - D_2$ & $-1$ & $(0,0)$ & $(0,0)$ & $+$ \\
$D_1 + D_2$ & $+1$ & $(0,0)$ & $(0,0)$ & $+$ \\
$D_3$ & $0$ & $(\frac12,0) \oplus (0,\frac12)$  & $(\frac12,\frac12)$ & $+$  \\
$D_4$ & $0$ & $(\frac12,0) \oplus (0,\frac12)$  & $(\frac12,\frac12)$ & $+$ \\
$D_5$ & $+1$ & $(1,0) \oplus (0,1)$ & $(1,0) \oplus (0,1)$ & $+$ \\
$D_6 - D_7$ & $-1$ & $(0,0)$ & $(1,0) \oplus (0,1)$ & $-$ \\
$D_6 + D_7$ & $+1$ & $(0,0)$ & $(1,0) \oplus (0,1)$ & $-$ \\
$D_8$ & $0$ & $(\frac12,0) \oplus (0,\frac12)$  & $(\frac12,\frac12)$ & $-$  \\
$D_9$ & $0$ & $(\frac12,0) \oplus (0,\frac12)$  & $(\frac12,\frac12)$ & $-$ \\
$D_{10}$ & $+1$ & $(1,0) \oplus (0,1)$ & $(0,0)$ & $-$ \\
 \hline
\end{tabular}
\label{tab:diquarks}
\end{center}
\end{table}
From the above Table \ref{tab:diquarks} we see that only the Lorentz
rep. $(\frac12,0) \oplus (0,\frac12)$ diquark fields have the same
chiral properties with both signs under the Pauli principle/two
particle exchange. This means that only the baryon fields made from
these diquarks are subject to ``Pauli mixing". These are the
$J=\half$ and $J=\thalf$ fields.

\section{Dirac, Isospin and Spatial Fierz Transformations}
\label{sec: app Fierz}

We summarize the detailed results of the Fierz transformation of
baryons. After the combined Fierz transformations of the iso-spin,
Dirac, spatial and color degrees of freedom, the Fierz transformed
field ${\cal F}[N]$ must satisfy the relation $N=-{\cal F}[N]$,
namely the Pauli principle.

In order to derive the following results, it is useful to have the
following Fierz identities.

\subsection{Isospin Fierz Transformations}
\label{sec: app Fierz1}

$\bullet$ Isospin:
\begin{eqnarray}
\left(\begin{array}{c} (\tau_0)_{12}(\tau_0)_{34} \\
(\tau_i)_{12}(\tau_i)_{34}\end{array}\right)
&=\half\left(\begin{array}{cc} 1 & 1 \\ 3 & -1 \end{array} \right)
\left(\begin{array}{c} (\tau_0)_{14}(\tau_0)_{23} \\
(\tau_i)_{14}(\tau_i)_{23}
\end{array}\right) \label{eq:20feb2008eq1}\\
(\tau^j)_{12}(P^{ij}_{3/2})_{34}&=(\tau^j)_{14}(P^{ij}_{3/2})_{32}.
\label{eq:3march2008eq1}
\end{eqnarray}

\subsection{Dirac Fierz Transformations}
\label{sec: app Fierz2}

$\bullet$ Spin:
\begin{eqnarray}
\left(\begin{array}{c} (1)_{12}(1)_{34} \\ (\gamma_5)_{12}(\gamma_5)_{34} \\
(\gamma^\mu)_{12}(\gamma_\mu)_{34}\\ (\gamma^\mu\gamma_5)_{12}(\gamma_\mu\gamma_5)_{34}\\
(\sigma^{\mu\nu})_{12}(\sigma_{\mu\nu})_{34}\end{array}\right)
& =\frac{1}{4}\left(\begin{array}{ccccc} 1 & 1 & 1 & -1 & \half \\
1 & 1 & -1 & 1 & \half\\
4 & -4 & -2 & -2 & 0 \\
-4 & 4 & -2 & -2 & 0 \\
12 & 12 & 0 & 0 & -2
\end{array}\right)
\left(\begin{array}{c} (1)_{14}(1)_{32} \\ (\gamma_5)_{14}(\gamma_5)_{32} \\
(\gamma^\mu)_{14}(\gamma_\mu)_{32}\\ (\gamma^\mu\gamma_5)_{14}(\gamma_\mu\gamma_5)_{32}\\
(\sigma^{\mu\nu})_{14}(\sigma_{\mu\nu})_{32}\end{array}
\right), \label{eq:20feb2008eq2}\\
\left(\begin{array}{c}
(\gamma_\nu)_{12}(\Gamma^{\mu\nu}_{3/2})_{34} \\
(\gamma_\nu\gamma_5)_{12}(\Gamma^{\mu\nu}_{3/2}\gamma_5)_{34} \\
i(\sigma_{\beta\alpha})_{12}(\Gamma^{\mu\beta}_{3/2}\gamma^\alpha)_{34}\end{array}\right)
&=
\frac{1}{2}\left(\begin{array}{ccc} 1 & 1 & -1 \\
1 & 1 & 1 \\ -2 & 2 & 0 \end{array}\right) \left(\begin{array}{c}
(\gamma_\nu)_{14}(\Gamma^{\mu\nu}_{3/2})_{32} \\
(\gamma_\nu\gamma_5)_{14}(\Gamma^{\mu\nu}_{3/2}\gamma_5)_{32} \\
i(\sigma_{\beta\alpha})_{14}(\Gamma^{\mu\beta}_{3/2}\gamma^\alpha)_{32}\end{array}\right)
\label{eq:3march2008eq2} \\
i(\sigma_{\beta\alpha})_{12}(\Gamma^{\mu\beta}_{3/2}\gamma^\alpha)_{34}
&= - i(\sigma_{\beta\alpha}
\gamma_5)_{12}(\Gamma^{\mu\beta}_{3/2}\gamma^\alpha \gamma_5)_{34}\\
(\sigma_{\alpha\beta})_{12}(\Gamma^{\mu\nu\alpha\beta}_{3/2})_{34}
&=
(\sigma_{\alpha\beta}\gamma_5)_{12}(\Gamma^{\mu\nu\alpha\beta}_{3/2}\gamma_5)_{34}
=(\sigma_{\alpha\beta})_{14}(\Gamma^{\mu\nu\alpha\beta}_{3/2})_{32}.
\label{eq:3march2008eq3}
\end{eqnarray}
Equations (\ref{eq:20feb2008eq1}) and (\ref{eq:20feb2008eq2}) are
well-known Fierz transformations, which are used to obtain
spin-$\half$ and isospin-$\half$ part of the Fierz identities, the
remaining identities Eqs. (\ref{eq:3march2008eq1}),
(\ref{eq:3march2008eq2}), (\ref{eq:3march2008eq3}) are necessary to
obtain the Fierz transformation of the baryons having spin-$\thalf$
and/or isospin-$\thalf$.

\begin{eqnarray}
\left(\begin{array}{c}
(\gamma_\nu)_{12}(\Gamma^{\mu\nu}_{3/2} \gamma_5)_{34} \\
(\gamma_\nu\gamma_5)_{12}(\Gamma^{\mu\nu}_{3/2})_{34} \\
i(\sigma_{\beta\alpha})_{12}(\Gamma^{\mu\beta}_{3/2}\gamma^\alpha
\gamma_5)_{34}\end{array}\right) &=
\frac{1}{2}\left(\begin{array}{ccc} 1 & 1 & 1 \\
1 & 1 & - 1 \\ 2 & - 2 & 0 \end{array}\right) \left(\begin{array}{c}
(\gamma_\nu)_{14}(\Gamma^{\mu\nu}_{3/2}\gamma_5)_{32} \\
(\gamma_\nu\gamma_5)_{14}(\Gamma^{\mu\nu}_{3/2})_{32} \\
i(\sigma_{\beta\alpha})_{14}(\Gamma^{\mu\beta}_{3/2}\gamma^\alpha
\gamma_5)_{32}\end{array}\right) \label{eq:25july2008eq2}
\end{eqnarray}

\subsection{Spatial Fierz transformation}
\label{sec: app spatial Fierz}

Let $P_{ij}$ be the $ij$-th particle interchange operator and
$(\vec{\rho}, \vec{\lambda})$ the three-body Jacobi coordinates
\begin{eqnarray}
\vec{\rho} &=& \frac{1}{\sqrt{2}}({\bf x_1} - {\bf x_2}),
\label{e:rho} \\ 
\vec{\lambda} &=& \frac{1}{\sqrt{6}}({\bf x_1} + {\bf x_2}- 2 {\bf
x_3}). \label{e:lambda} \
\end{eqnarray}
Then the three-particle permutation/exchange symmetry $s_3$ can be
examined as
\begin{eqnarray}
P_{12}\vec{\rho} & \rightarrow -\vec{\rho} \\
P_{12}\vec{\lambda} & \rightarrow \vec{\lambda} \\
P_{13}\vec{\rho} & \rightarrow \frac{1}{2}\vec{\rho} -
\frac{\sqrt{3}}{2}\vec{\lambda} \\
P_{13}\vec{\lambda} & \rightarrow -\frac{\sqrt{3}}{2}\vec{\rho} -
\frac{1}{2}\vec{\lambda}
\end{eqnarray}
i.e.,
\begin{eqnarray}
{\cal F}\left[\left(
\begin{array}{c} \vec{\rho} \\
\vec{\lambda} \end{array}\right)\right] &=& P_{23} \left[\left(
\begin{array}{c} \vec{\rho} \\
\vec{\lambda} \end{array}\right)\right] = \frac12 \left(
\begin{array}{cc} 1 & \sqrt{3} \\ \sqrt{3} & -1 \end{array}\right)
\left(\begin{array}{c} \vec{\rho} \\
\vec{\lambda} \end{array}\right) .
\end{eqnarray}

%
%

\end{document}